\documentclass[preprints,review,accept,moreauthors,pdftex]{Definitions/mdpi}

\firstpage{1} 
\pubvolume{1}
\issuenum{1}
\articlenumber{0}
\pubyear{2021}
\copyrightyear{2020}
\datereceived{} 
\dateaccepted{} 
\datepublished{} 
\hreflink{https://doi.org/} 
\pdfoutput=1

\usepackage{subfigure}

\Title{A historical review and Bibliometric analysis of research on Weak measurement research over the past decades based on Biblioshiny}

\TitleCitation{ A historical review and Bibliometric analysis of research on Weak measurement research over the past decades based on Biblioshiny}

\Author{Jing-Hui Huang $^{1,\dagger,\ddagger}$\orcidA{}, Xue-Ying Duan $^{2,3,4\ddagger}$\orcidB{}, Fei-Fan He$^{1,\ddagger}$\orcidC{}, Guang-Jun Wang $^{2,3,4\ddagger}$\orcidD{} and Xiang-Yun Hu $^{1,}$*\orcidE{}}

\AuthorNames{Jing-Hui, Huang; Xue-Ying Duan; Fei-Fan He; Guang-Jun Wang and Xiang-Yun Hu}

\AuthorCitation{Jing-Hui, Huang; Xue-Ying Duan; Fei-Fan He; Guang-Jun Wang and Xiang-Yun Hu}

\address{%
$^{1}$ \quad School of Institute of Geophysics and Geomatics, China University of Geosciences, Lumo Road 388, 430074 Wuhan, China.\\
$^{2}$ \quad School of Automation, China University of Geosciences, Lumo Road 388, 430074 Wuhan, China.\\
$^{3}$ \quad Hubei Key Laboratory of Advanced Control and Intelligent Automation for Complex Systems, China\\
$^{4}$ \quad Engineering Research Center of Intelligent Technology for Geo-Exploration, Ministry of Education, China}

\corres{Correspondence: xyhu@cug.edu.cn)}

\secondnote{These authors contributed equally to this work.}
\abstract{Weak measurement has enabled fundamental studies in both experiment and theory of quantum measurement. Numerous researches have indicated that weak measurements have a wide range of application and scientific research value. In our work, we used bibliometric methods to evaluate the global scientific output of research on Weak measurement and explore the current status and trends in this field from 2000 to 2020. The R bibliometric package was used for quantitative and qualitative analyses of publication outputs and author contributions. In total, 636 related articles and reviews were included in the final analysis. The main results were as follows: The number of publications has increased substantially with time. Physical Review A was the most active journal. The country and institution contributing the most to this field were The United States and University Rochester respectively.  The analysis of the conceptual structure of keywords indicated that weak measurements were involved a variety of metrology, quantum communication, and nonclassical features of quantum mechanics. Our bibliometric analysis shows that weak measurement research continues to be a hot-spot. The focus has evolved to study quantum information and amplify weak signals.}


\begin{document}
\section{Introduction}
\label{intro}
Recently, weak measurement has become an intensive and important area of research\cite{RevModPhys.86.307,2011Nonperturbative,2012Protecting,e23030354} in quantum mechanics. In quantum weak measurement, the weak value, obtained from a weak measurement followed by a strong measurement, is quite peculiar in that it is, in general, a complex number and is not bounded by the eigenvalue spectrum of the associated observable\cite{AAV,2018Direct}. The "Weak-value" was first proposed by Aharonov et al\cite{AAV} as “a new kind of value for a quantum variable”, where information is gained by weakly coupling the probe to the system. 

Unlike the conventional/projection measurement, where a quantum system is irrecoverably collapsed into one of the eigenstates of the observable, the weak measurement can break this constraint and has enabled novel research into important problems in quantum physics and quantum information,  e.g., protecting quantum states from decoherence\cite{2012Protecting,Kim:09}, direct measurement of the quantum wavefunction\cite{0Directnature}, direct observation of geometric phases related to the three-vertex Bargmann invariant\cite{PhysRevA.81.012104}, direct observation of Hardy's paradox by joint weak measurement\cite{Yokota_2009}, observing the direct path state characterization by strongly measuring weak values\cite{PhysRevLett.118.010402}, experimental realization of the quantum three-box problem\cite{RESCH2004125}, observation of a quantum Cheshire Cat in a matter-wave interferometer experiment\cite{2014Observation}. 
 
In addition, the weak value amplification(WVA) technique can help to amplify a detector signal and enable the sensitive estimation of unknown small evolution parameters. Therefore, weak measurement has also been utilized in metrology\cite{PhysRevA.92.032127,Fang_2021,Huang2021,PhysRevA.102.042601,PhysRevA.103.053518}, and it has a variety of applications in precision detection as well as its advantage of high precision\cite{e23030354}. Specifically, the Spin Hall Effect of Light can be enhanced by nearly four orders of magnitude with the WVA technique\cite{Hosten787}; an interferometric weak value technique can amplify very small transverse deflections of an optical beam\cite{PhysRevLett.102.173601}; a weak measurement protocol can permit a sensitive estimation of angular rotations in the azimuthal degree of freedom \cite{PhysRevLett.112.200401}. Note that researchers all over the world have studied the trends of weak measurement research and published many papers related to the topic. However, a comprehensive statistical review of the global estuary pollution research has never been done.

Bibliometric, which was firstly introduced by Pritchard, is the quantitative and qualitative analysis of published academic literature for tracking the development of a certain research field over a long period\cite{1969Statistical,2003Paradigms}. Bibliometrics can demonstrate information of a certain field to researchers by investigating the publication characteristics, such as authorship, sources, institution, journals, citations, corresponding author's country, and even co-citation Network\cite{2003Paradigms}. Recently, Bibliometrix has wide applications in various fields to elevate research performance or assess the research trends, such as, Celik et al. used Bibliometrix analysis to integrate worldwide studies of Hg in soil that were published between 1991 and 2020\cite{2021Celik}. Guo et al. used bibliometric methods to evaluate the global scientific output of research on Piezo channels and explore the current status and trends in this field over the past decades\cite{2021Trends}. Lyu et al. introduced a novel bibliometric approach to unfold the status of a given scientific community from an individual-level perspective\cite{2021Studying}.

\begin{figure*}[htbp]
\vspace*{-5mm}
\centering \includegraphics[width=0.96\textwidth]{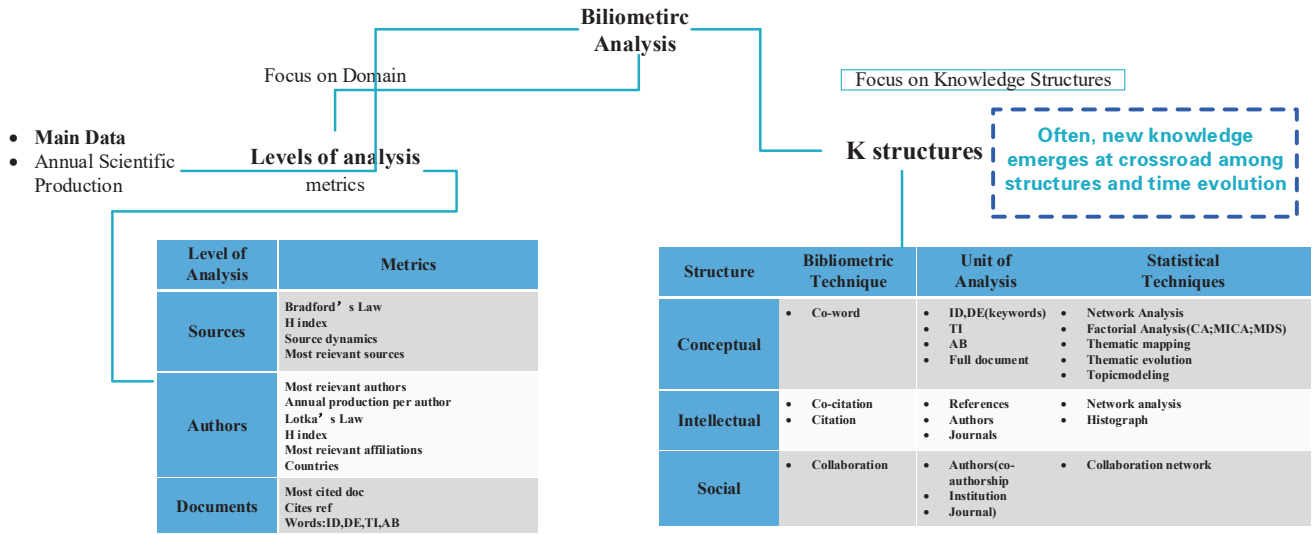}
\vspace*{-3mm}
\caption{Bibliometric analysis for systematic literature reviews.(Adapted from Ref. \cite{ARIA2017959})}
\label{Fig:workflow_bibiometrix}
\end{figure*}

In this study, we aimed to provide a report on scientific publication in research on weak measurement over the past decades(2000-2020). In section II, we give a brief introduction to the  Biblioshiny Bibliometrix R-package\cite{biblitoolbox-web}, which allows no codes to use bibliometric and enables visualization in this study. Then the results including general information, citation and H-index analysis, active authors, active journals, and analysis of keywords are shown in section III. Sec IV is devoted to the summary and the discussions.

\section{Data sources and methods }
\label{sec:Data_sources}

The data used in this study were retrieved from the Science Citation Index-Expanded (SCI-E) of the Web of Science (WOS) Core Collection on July 20, 2021. The search terms were "weak measurement", "weak value" and "postselection" with publication timespan(2000–2020)  to compile a bibliography of all articles related to the research on weak measurement research. In total, 636 publications were obtained and analyzed by choosing the following publishers: "Amer Physical Soc", "Springer Nature", "Top Publishing LTD", "Elsevier", "Optical Soc Amer" and  "Amer inst Physics". At last, these data sets were converted to text format and imported to the bibliophily for Bibliometrix in R 3.6.3 and included the distribution of countries/regions, years of publication, and authors.

Biblioshiny for bibliometrix\cite{biblitoolbox-web,ARIA2017959} is a Java software developed by Massimo Aria from the University of Naples Federico. Biblioshiny combines the functionality of the bibliometrix package with the ease of use of web apps using the Shiny package environment. The main systematic literature reviews of the bibliophily for bibliometric analysis are shown in Fig. \ref{Fig:workflow_bibiometrix}. Normally, the characteristics of publications, including authors, journals, institutions, citations, H-index, M-index and the corresponding author's country can be reviewed by establishing the WOS literature Analysis Report online. Where "H-index" is used to quantify an individual's scientific research output and measure his citation impact\cite{2005An}.  The M-index was proposed to facilitate comparisons between academics with different academic careers lengths M-index=H-index/N (N is measured as the number of years since the first published paper in the research area)\cite{2021Trends}. In addition, the index can be used as a tool in predicting future research by analyzing the trend topics of keywords.

However, the Bibliometrix package can provide various routines for importing bibliographic data, performing bibliometric analysis, and building data matrices for co-citation, coupling, scientific collaboration analysis, and co-word analysis. In addition, new knowledge always emerges at crossroads among structures and time evolution, such as network analysis, factorial analysis, thematic mapping. Note that these results can be visualized by using the bibliophily app. And the results will be shown in the next section.

\section{Results}

\subsection{General information and annual publication output}
\begin{figure*}[htbp]
	\centering
	\centerline{\includegraphics[scale=0.52,angle=0]{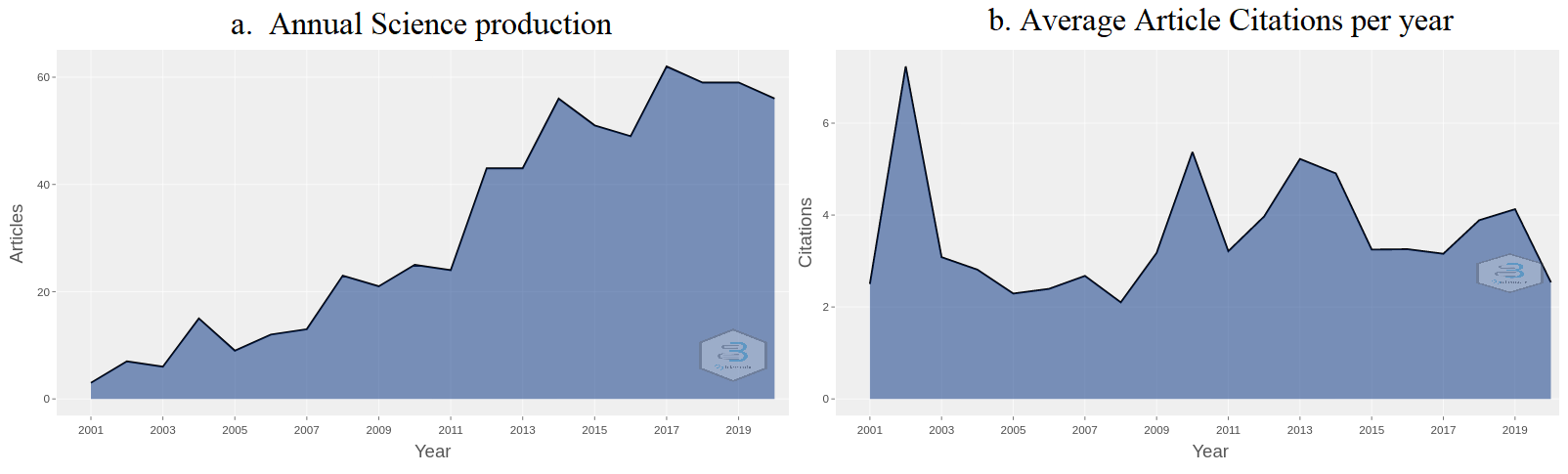}}
\vspace*{-3mm} 
\caption{\label{Fig:result-Annual-pubulation}(a) Global trends in publications on weak measurement from 2000 to 2020; and (b) average article citations per year. 
}
\end{figure*}

\begin{figure*}[htbp]
	\centering
	\centerline{\includegraphics[scale=0.52,angle=0]{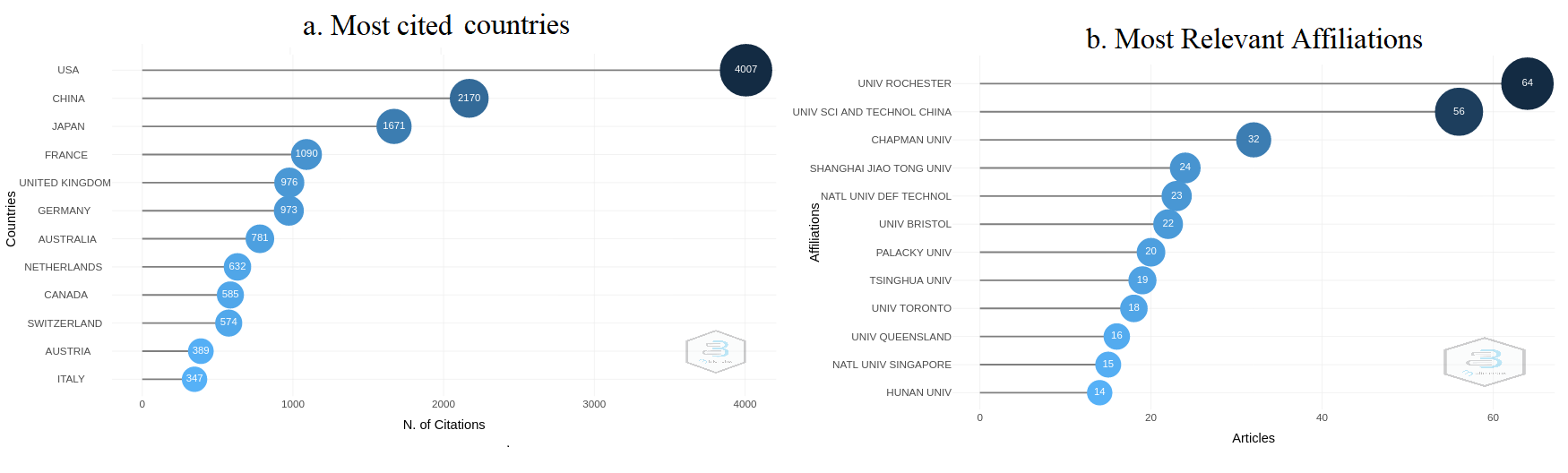}}
\vspace*{-6mm} 
\vspace*{0mm} 
\caption{\label{Fig:Most-cited-countries}(a) Global trends in publications on weak measurement from 2000 to 2020; and (b) average article citations per year. 
}
\end{figure*}

\begin{figure*}[htbp]
	\centering
\subfigure
{
	\vspace{-0.2cm}
	\begin{minipage}{9.4cm}
	\centering
	\centerline{\includegraphics[scale=0.6,angle=0]{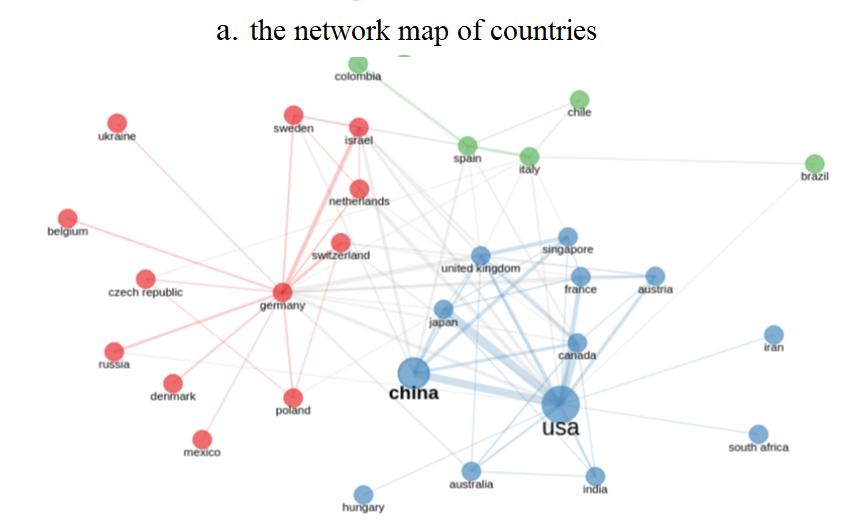}}
	\end{minipage}
}

\subfigure
{
	\vspace{-0.2cm}
	\begin{minipage}{9.4cm}
	\centering
	\centerline{\includegraphics[scale=0.6,angle=0]{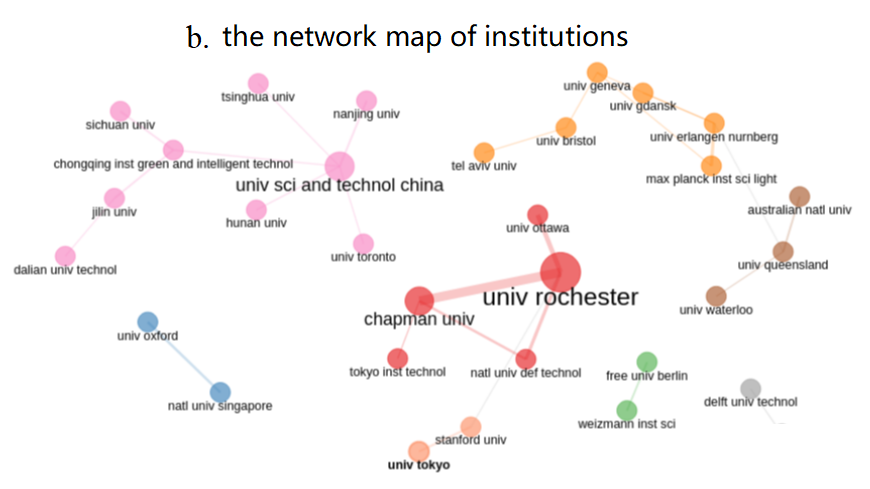}}
	\end{minipage}
}
\vspace*{-4mm} 
\vspace*{0mm} \caption{\label{Fig:networkcountry}Collaboration Network analysis of countries and institutions. (a) the network map of institutions, and (b) the network map of countries.
}
\end{figure*}

\begin{figure*}[htbp]
	\centering
\subfigure
{
	\vspace{-0.2cm}
	\begin{minipage}{9.4cm}
	\centering
	\centerline{\includegraphics[scale=0.52,angle=0]{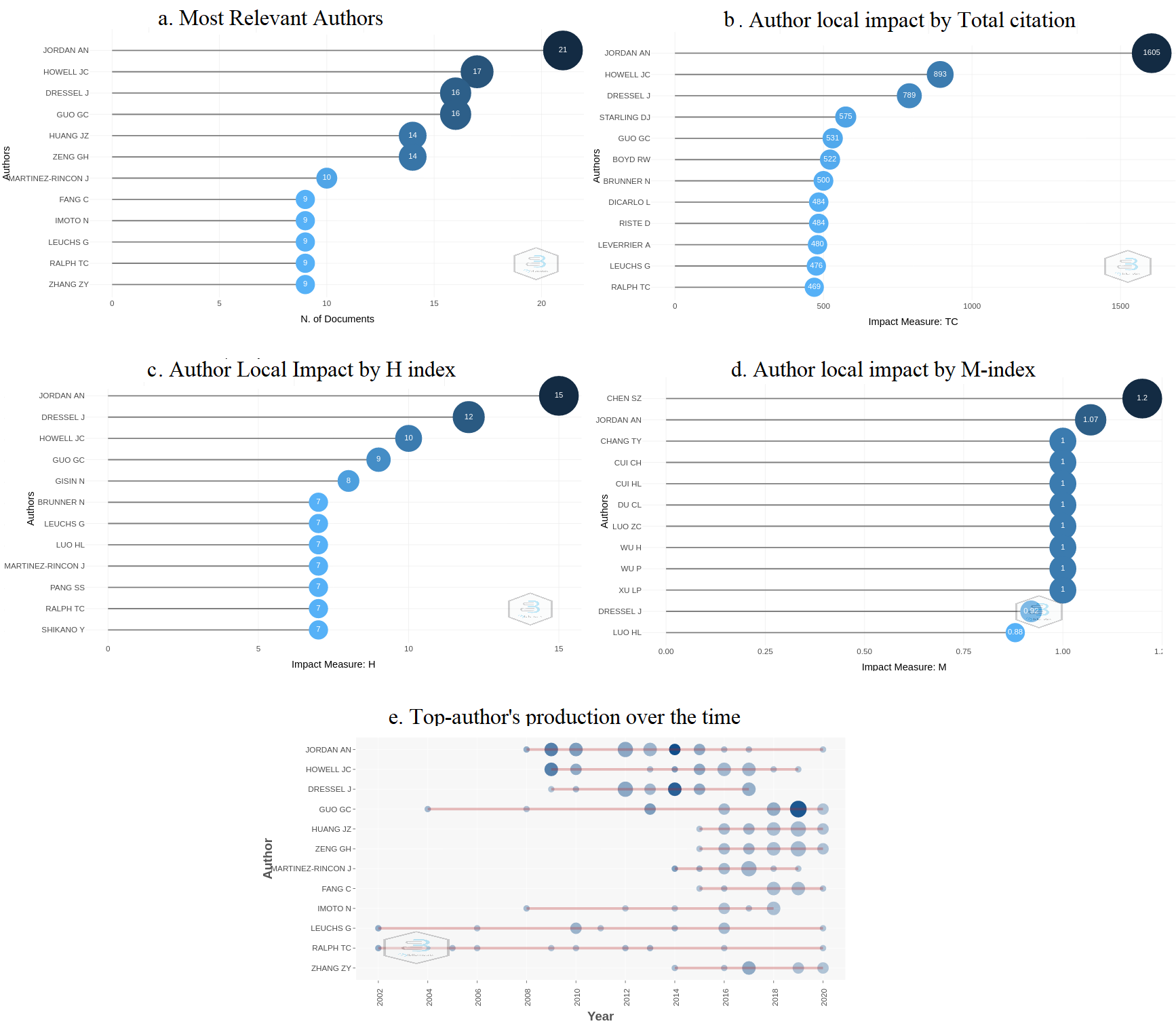}}
	\end{minipage}
}
\vspace*{-6mm} 
\vspace*{0mm} \caption{\label{Fig:mostrelevantAuthors}Analysis of the active author: (a) The top 12 most relevant authors; (b)Total citations in the research filed from different authors; (c)H-index of publications from different authors; (d) and M-index of publications from different authors;(e) Top-author's production from 2000 to 2020.
}
\end{figure*}

\begin{figure*}[htbp]
	\centering
\subfigure
{
	\vspace{-0.2cm}
	\begin{minipage}{9.4cm}
	\centering
	\centerline{\includegraphics[scale=0.56,angle=0]{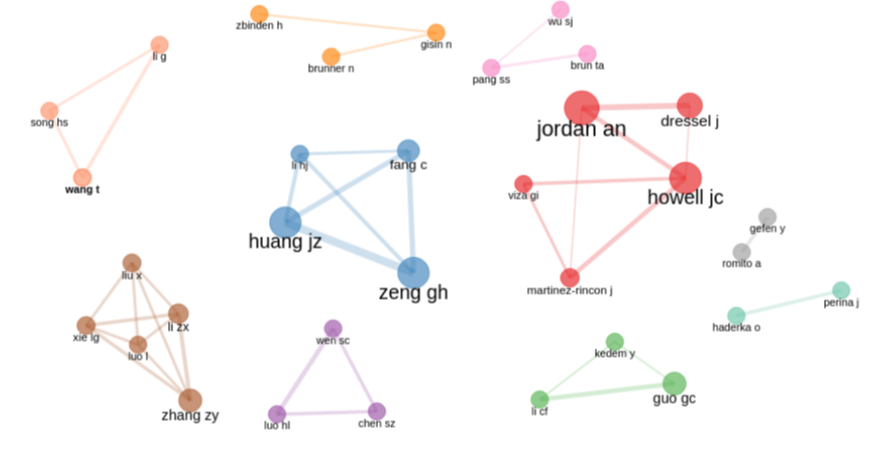}}
	\end{minipage}
}
\vspace*{-3mm} 
\vspace*{0mm} \caption{\label{Fig:network-author}Network map of co-occurrence between authors on the weak measurement research.
}
\end{figure*}

\begin{figure*}[htbp]
	\centering
\subfigure
{
	\vspace{-0.2cm}
	\begin{minipage}{9.4cm}
	\centering
	\centerline{\includegraphics[scale=0.52,angle=0]{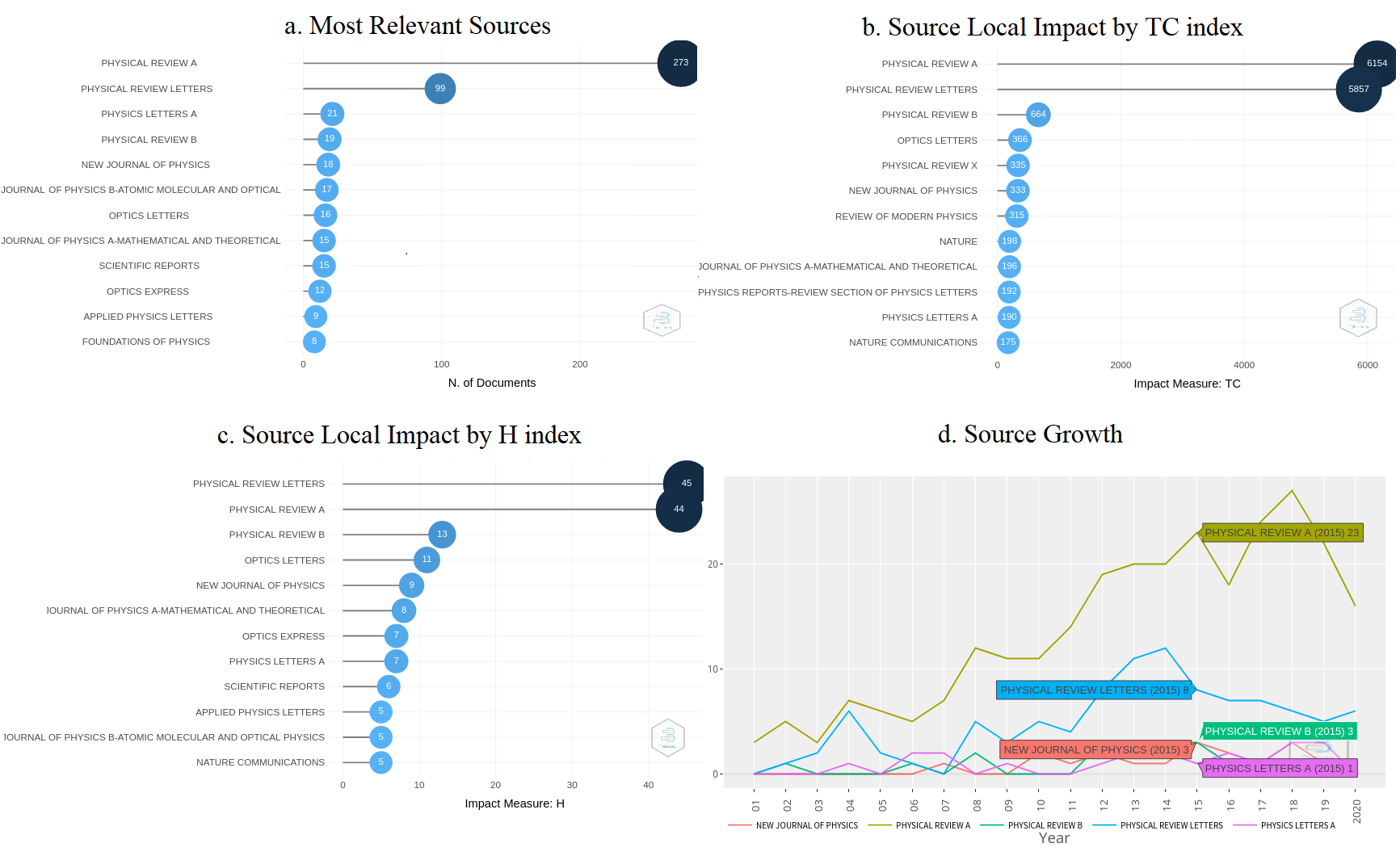}}
	\end{minipage}
}
\vspace*{-3mm} 
\vspace*{0mm} \caption{\label{MostRelevantSources}Analysis of sources. (a) Most relevant sources; (b) Source local impact by total citations(TC) index; (c)Author local impact by H-index; (d) Source Growth on weak measurement research from 2000 to 2020. 
}
\end{figure*}

\begin{figure*}[htbp]
	\centering
\subfigure
{
	\vspace{-0.2cm}
	\begin{minipage}{9.4cm}
	\centering
	\centerline{\includegraphics[scale=0.52,angle=0]{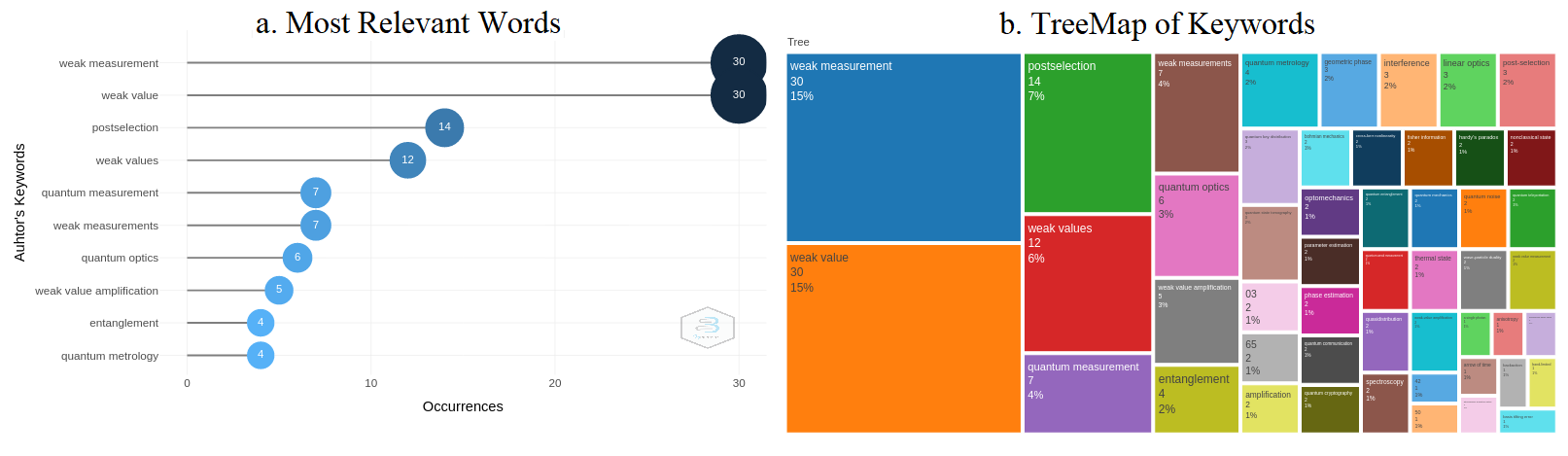}}
	\end{minipage}
}
\vspace*{-3mm} 
\vspace*{0mm} \caption{\label{Fig:MostrelavantWorlds}Analysis and Co-occurrence Network of Keywords on weak measurement research. (a) Most relevant keywords; (b)TreeMap of Keywords on weak measurement research from 2000 to 2020. 
}
\end{figure*}
\begin{figure*}[htbp]
	\centering
\subfigure
{
	\vspace{-0.2cm}
	\begin{minipage}{9.4cm}
	\centering
	\centerline{\includegraphics[scale=0.63,angle=0]{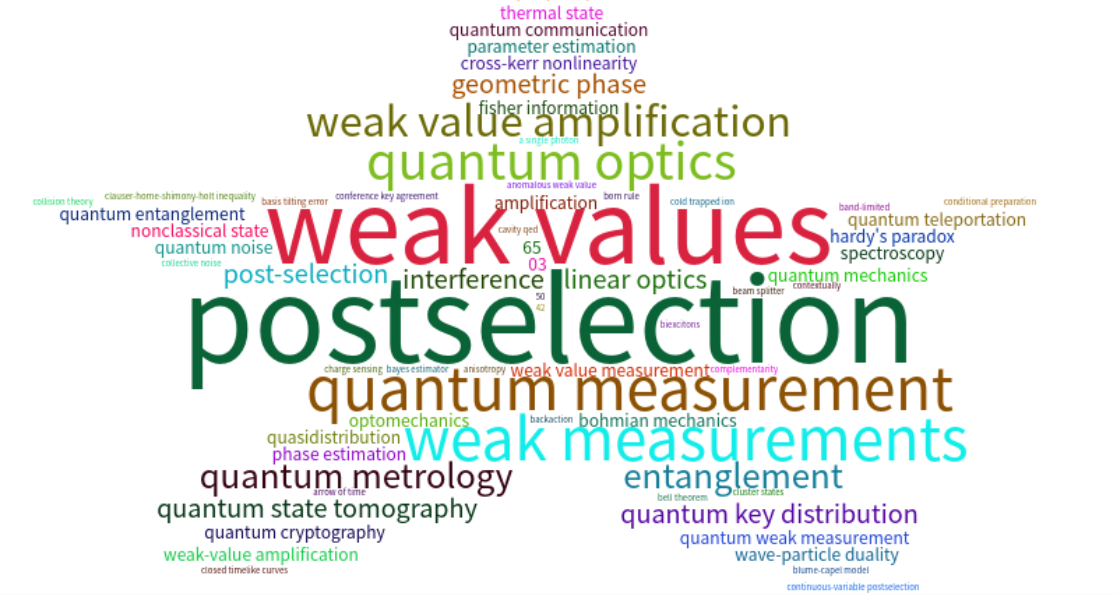}}
	\end{minipage}
}
\vspace*{-5mm} 
\vspace*{0mm} \caption{\label{Fig:WordsCloud}Keywords cloud on weak measurement research from 2000 to 2020. 
}
\end{figure*}
\begin{figure}[htbp]
	\centering
\subfigure
{
	\vspace{-0.2cm}
	\begin{minipage}{9.4cm}
	\centering
	\centerline{\includegraphics[scale=0.63,angle=0]{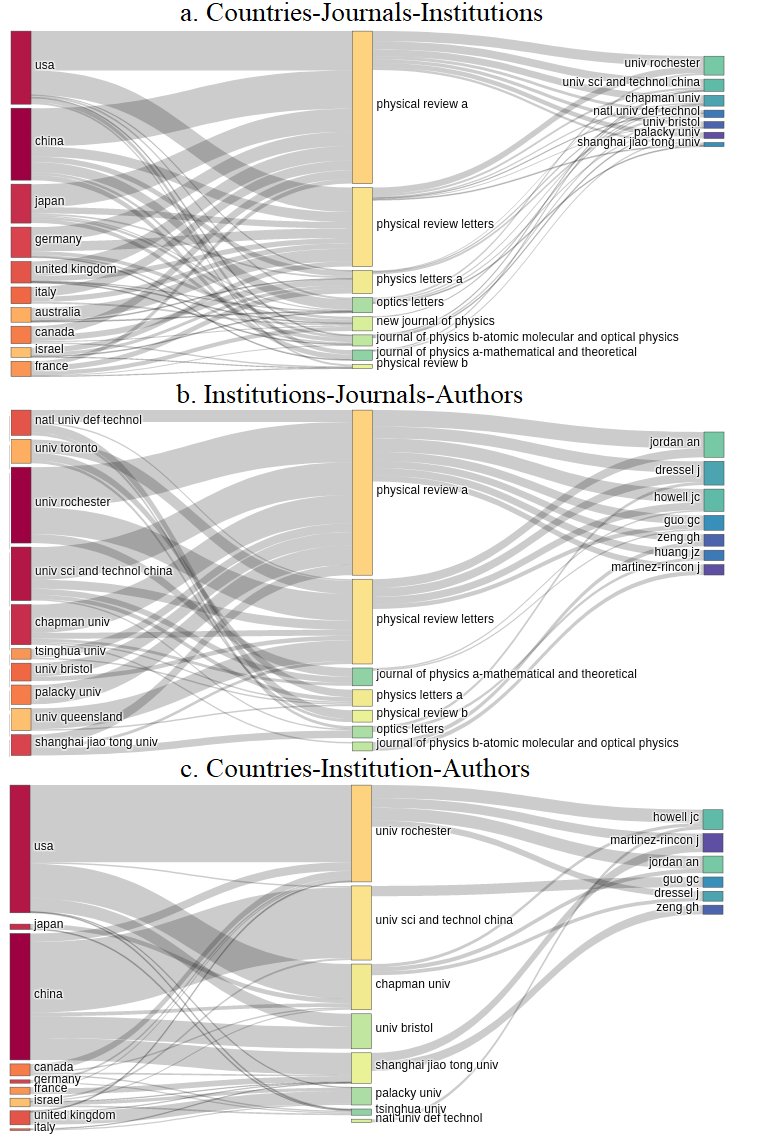}}
	\end{minipage}
}
\vspace*{-5mm} 
\vspace*{0mm} \caption{\label{Fig:institution-source-authour}Three-Fields Plot among countries, institutions, journals and authors on weak measurement research.
}
\end{figure}
\begin{figure*}[htbp]
	\centering
\subfigure
{
	\vspace{-0.2cm}
	\begin{minipage}{9.4cm}
	\centering
	\centerline{\includegraphics[scale=0.7,angle=0]{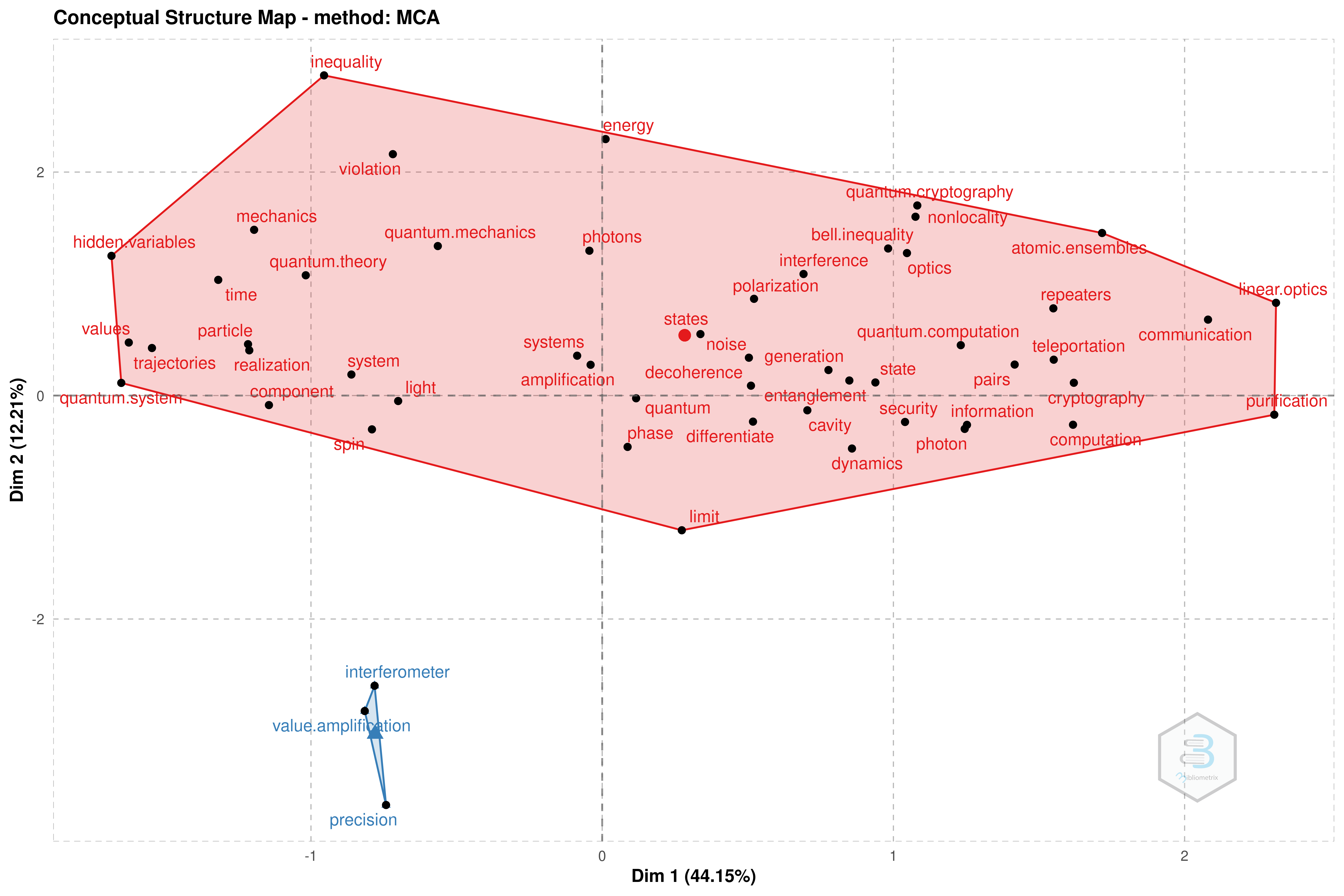}}
	\end{minipage}
}
\vspace*{-5mm} 
\vspace*{0mm} \caption{\label{Fig:MCA}Conceptual structure map with Multiple Correspondence Analysis of Keywords on weak measurement research.
}
\end{figure*}
\begin{figure*}[htbp]
	\centering
\subfigure
{
	\vspace{-0.2cm}
	\begin{minipage}{9.4cm}
	\centering
	\centerline{\includegraphics[scale=0.42,angle=0]{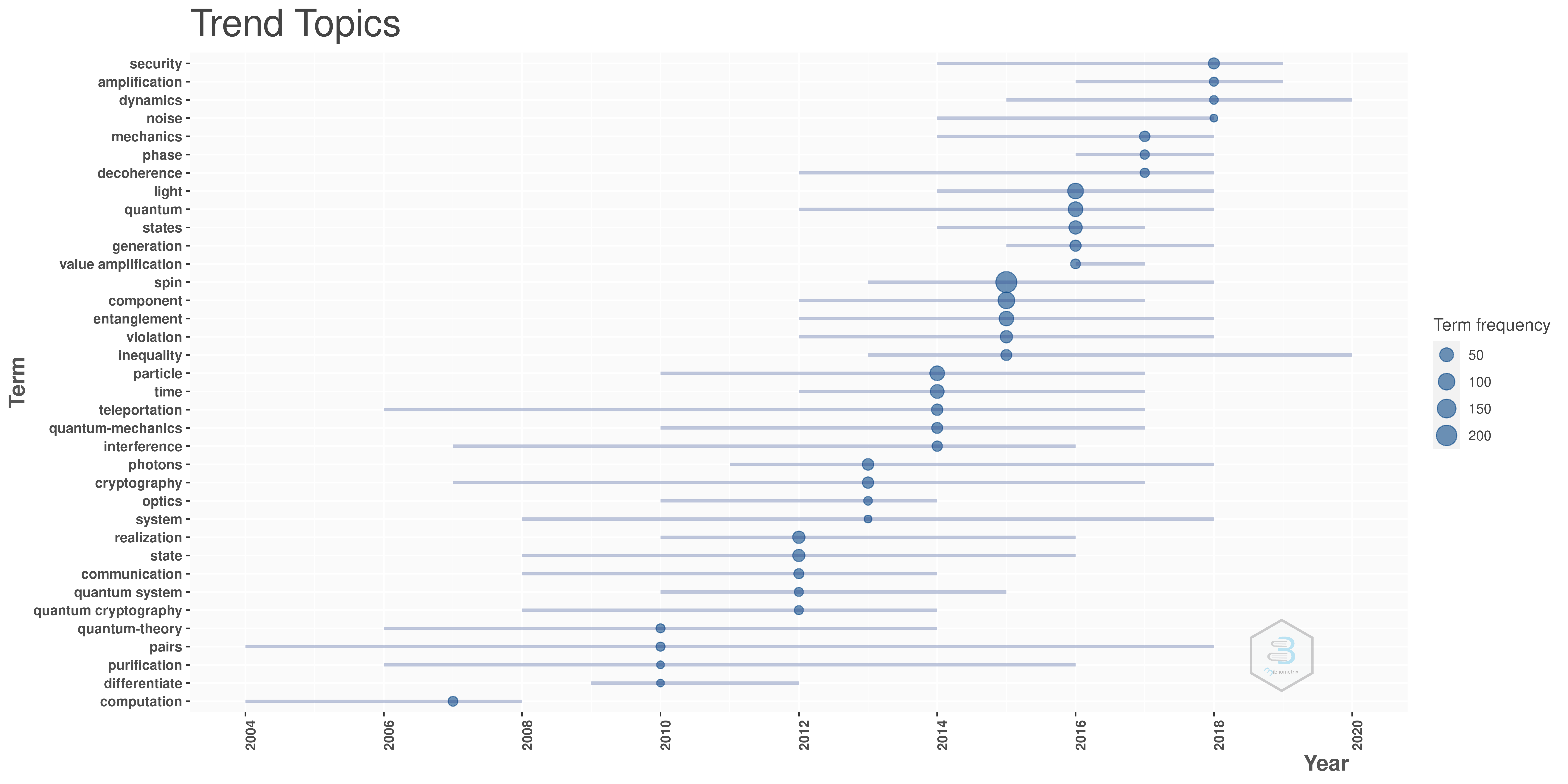}}
	\end{minipage}
}
\vspace*{-5mm} 
\vspace*{0mm} \caption{\label{Fig:trendtopics}Trend Topics of Keywords on weak measurement research from 2000 to 2020.
}
\end{figure*}

\begin{table}[htbp]
\centering
\caption{Main information about the data and the document types from the WOS.}
\label{tab:Mian-information}   
\begin{tabular}{p{6.5cm}p{1.5cm}}
\toprule
Dsecription & Results\\
\noalign{\smallskip}\hline\noalign{\smallskip}
Mian information about the data & \\
\noalign{\smallskip}\hline\noalign{\smallskip}
Timespan                        & 2001:2020\\
Source(Journals,Book,etc)         & 61\\
Documents                       & 636\\
Average years from publication    & 6.92\\
Average citations per documents    & 25.52\\
Average citations per year per doc    & 3.021\\
\noalign{\smallskip}\hline\noalign{\smallskip}
Document types    & \\
\noalign{\smallskip}\hline\noalign{\smallskip}
Article        & 628\\
Proceedings paper & 5\\
Review & 2\\
\toprule\\
\end{tabular}
\vspace*{-0.4cm}  
\end{table}

Table. \ref{tab:Mian-information} shows the main information about the data and the document types from the WOS. A total of 636 articles from 2000 to 2020 and related to weak measurement were retrieved from WOS. We presented the number of articles per year in Fig. \ref{Fig:result-Annual-pubulation}a, which showed that the number of publications related to the weak measurement research has generally increased. Moreover, the overall trend increased from three papers in 2001 to 56 papers in 2020. And the annual growth rate reaches 16.65$\%$. In addition, the number of the average article citations per year in Fig. \ref{Fig:result-Annual-pubulation}b shows that the average number of citations for weak measurement studies can remain stable. Therefore, we can conclude that the field is currently in a phase of steady growth in global trends in publications.

\subsection{Distribution of Institutions and Countries}
\begin{table}[htbp]
\centering
\caption{The top 10 countries contributed to the total citations on weak measurement research. Average$^1$ represents the average article citations.}
\label{tab:Most-cited-country}   
\begin{tabular}{p{1.1cm}p{3.1cm}p{2.4cm}p{1.2cm}}
\toprule
Rank & Country &Total Citations& Average$^1$ \\
\noalign{\smallskip}\hline\noalign{\smallskip}
1 & USA & 4007 & 39.28\\
2 & China & 2170 & 17.5\\
3 & Japan & 1671 & 23.87\\
4 & France & 1090 & 57.37\\
5 & United Kingdom & 976 & 25.68\\
6 & Germany & 973 & 31.39\\
7 & Australia & 781 & 32.54\\
8 & Netherlands & 632 & 52.67\\
9 & Canada & 585 & 27.86\\
10 &Switzerland & 574 & 47.83\\
\toprule\\
\end{tabular}
\vspace*{-0.4cm}  
\end{table}
\begin{table}[!htbp]
\centering
\caption{The top 10 institutions contributed to the total publications on weak measurement research. USTC$^1$: University of Science and Technology of China; SJTU$^2$: Shanghai Jiao Tong University; NUODT$^3$: National University of Defense Technology}
\label{tab:Most-cited-affiliation}   
\begin{tabular}{p{1.1cm}p{3.6cm}p{1.4cm}p{1.6cm}}
\toprule
Rank & Institutions &Articles& Proportion \\
\noalign{\smallskip}\hline\noalign{\smallskip}
1 & University Rochester & 64 & 10.0$\%$\\
2 & USTC$^1$ & 56 & 8.8$\%$\\
3 & Chapman University & 32 & 5.0$\%$\\
4 & SJTU$^2$ & 24 &3.7$\%$\\
5 & NUODT$^3$ & 23 & 3.6$\%$\\
6 & University Bristol & 22 &  3.4$\%$\\
7 & Palacky University & 20 & 3.1$\%$\\
8 & Tsinghua University & 19 & 2.9$\%$\\
9 & University Toronto & 18 & 2.8$\%$\\
10 &University Queensland & 16 & 2.5$\%$\\
\toprule\\
\end{tabular}
\vspace*{-0.4cm}  
\end{table}
Global contributions to weak measurement research were analyzed and represented in Fig.\ref{Fig:Most-cited-countries}. And the contribution to the total citation from different countries is shown in Table. \ref{tab:Most-cited-country}. A total of 36 countries participated in a weak measurement study. By counting the number of the corresponding author's country, the most cited countries in this field are shown in  Fig. \ref{Fig:Most-cited-countries}a. The United States contributed the greatest number of citations (4007, 25.7$\%$ of all citations), followed by China (2170, 13.9$\%$) and Japan(1671, 10.7$\%$). The cooperative relationship among these countries is demonstrated in Fig .\ref{Fig:networkcountry}a. The thickness of the lines indicates the strength of the relationship. The United state(USA) and China show equal importance to global cooperation. Meanwhile, the USA and China had close collaborations with each other.

Note that the order of average article citations is not consistent with the order of total Citations in Table. \ref{tab:Most-cited-country}. In the top 10 countries ranked by total citations, articles from the Netherlands had the highest number of the average article citations. In other words, the average quality of the articles from the Netherlands is the best. 

A total of 186 institutions were involved in weak measurement research. Most of the articles originated from affiliations in the USA, with University Rochester producing the highest number of articles(64 records, 10.0$\%$) on weak measurement, followed by the University of Science and Technology of China(56 records, 8.8$\%$) and Chapman University(32 records, 3.6$\%$). In addition, the connection among these affiliations is demonstrated in Fig. \ref{Fig:Most-cited-countries}b.

\subsection{Analysis Authors}
\begin{table}[htbp]
\centering
\caption{The top 10 most active authors on weak measurement research.}
\label{tab:Most-cited-authors}   
\begin{tabular}{p{1.2cm}p{3.1cm}p{1.2cm}p{2.3cm}}
\toprule
Rank & Author &Articles&  Fractionalized \\
\noalign{\smallskip}\hline\noalign{\smallskip}
1 & Andrew N. Jordan   & 21 & 6.82\\
2 & John C. Howell     & 17 & 4.06\\
3 & Dressel, Justin    & 16 & 6.10\\
4 & Guang-Can Guo      & 16 &2.13\\
5 & Jing-Zheng Huang   & 14 & 3.18\\
6 & GH. Zeng           & 14 & 3.18\\
7 & J. Martinez-Rincon & 10 & 3.48\\
8 & Chen Fang          & 9  & 2.03\\
9 & Nobuyuki Imoto     & 9  & 4.03\\
10 &ZY. Zhang          & 9  & 2.07\\
\toprule\\
\end{tabular}
\vspace*{-0.4cm}  
\end{table}
A total of 1452 authors contributed 636 publications related to weak measurement research. Fig. \ref{Fig:mostrelevantAuthors}a shows the top 12 most relevant authors. In terms of the number of publications is shown in Table. \ref{tab:Most-cited-authors}. Andrew N. Jordan was the most productive author, with 21 articles (3.3$\%$ of all articles), followed by John C. Howell(2.6$\%$ of all articles) and J. Dressel(2.5$\%$ of all articles). In terms of citations in this field, Andrew N. Jordan was ranked first (1605
citations), followed by John C. Howell (893 citations), Dressel, Justin (789 citations), DJ. Starling (575 citations), and Guang-Can Guo (531 citations). Note that from Fig. \ref{Fig:mostrelevantAuthors}a, Fig. \ref{Fig:mostrelevantAuthors}b and Fig. \ref{Fig:mostrelevantAuthors}c, Andrew N. Jordan was the most active author on the weak measurement research, and whose works laid the foundation for the major research projects being undertaking today. Fig. \ref{Fig:mostrelevantAuthors}d shows that SZ. Chen has an M-index of 1.2 more than Andrew N. Jordan, suggesting that he can be characterized as the most promising scientist.

In addition, Fig. \ref{Fig:mostrelevantAuthors}e demonstrates the Top-author's production from 2000 to 2020. We can conclude that most of the articles by top-author had been published in the last decade. On the other hand, the result shows that more and more outstanding scientists, such as Chen Fang and Jing-Zheng Huang have devoted themselves to weak measurement research. This means that the study of weak measurement will continue to be a hot field.

We analyzed a total of 33 authors that were co-authored in more than five publications and the network map of co-occurrence between authors on the weak measurement research is shown in Fig \ref{Fig:network-author}. Our findings suggest that authors' partnerships are largely confined to the country in which the authors reside. For example, Jing-Zheng Huang, Chen Fang, and GH. Zeng keep the closest cooperation with each other, and they belong to China.

\subsection{Active journals}
\begin{table}[htbp]
\centering
\caption{The top 12 journals ranked by Source local impact by total citation index. J. Phys. A: Journal Of Physics A-mathematical and Theoretical; Rev Mod Phys: Reviews Of Modern Physics.}
\label{tab:Source-local-impact}   
\begin{tabular}{p{1.1cm}p{3.6cm}p{1.4cm}p{1.6cm}}
\toprule
Rank & Journal & Citations& Impact Factor \\
\noalign{\smallskip}\hline\noalign{\smallskip}
1 & Physical Review A      & 6454 & 3.140\\
2 & Physical Review Letter & 5858 & 9.161\\
3 & Physical Review B      & 664  & 4.036\\
4 & Optics Letters         & 366  & 3.776\\
5 & Physical Review X      & 335  & 15.76\\
6 & New Journal Of Physics & 333  & 3.729\\
7 & Rev Mod Phys  & 315 & 54.49\\
8 & Nature & 198 & 49.96\\
9 & J. Phys. A & 196 & 2.868\\
10 &Physics Reports  & 192 & 25.60\\
\toprule\\
\end{tabular}
\vspace*{-0.4cm}  
\end{table}
The 636 articles were published in a total of 61 journals, with 17 journals have published more than five papers on weak measurement research. the top 12 journals in terms of the number of publications are shown in Fig. \ref{MostRelevantSources}a. The number 273 of articles published in “Physical Review A” is the largest, followed by “Physical Review Letter”(99 publications). 

In addition, the top 12 journals ranked by Source local impact by total citation index are shown in Fig. \ref{MostRelevantSources}b and Table \ref{tab:Source-local-impact}. Physical Review A(IF = 3.140) had 6454 total citations on the weak measurement research, which was the highest number, followed by Physical Review Letter(IF = 9.161). Note that these journals such as Nature(IF = 49.96) and Rev Mod Phys(IF = 54.49) had higher impact factors while had lower total citations due to fewer publications. The higher the impact factor, the more difficult it is to publish.

Fig. \ref{MostRelevantSources}c shows the top 12 journals ranked by Source local impact by H-index. Normally, H-index is used as a tool in predicting future research. Therefore, our results show that Physical Review A and Physical Review Letter were the most active journals with high citations and would keep the trends in the future. Meanwhile, we can obtain the same conclusion from Fig. \ref{MostRelevantSources}d.

\subsection{Analysis of high-citation documents}
\begin{table*}[htbp]
\centering
\caption{Top ten citation analysis of documents on weak measurement research.}
\label{tab:analysis-of-documents}   
\begin{tabular}{p{0.8cm}p{8.9cm}p{1.4cm}p{2.2cm}p{3.3cm}}
\toprule
Rank & Title & Citations& First Author &DOI \\
\noalign{\smallskip}\hline\noalign{\smallskip}
1 & Ultrasensitive Beam Deflection Measurement via Interferometric Weak Value Amplification\cite{PhysRevLett.102.173601}     & 412 &  P. Ben Dixon  &\url{10.1103/PhysRevLett.102.173601}\\
2 & Colloquium: Understanding quantum weak values: Basics and applications\cite{RevModPhys.86.307} & 315 &Justin Dressel & \url{10.1103/RevModPhys.86.307}\\
3 & Monoclinic structure of unpoled morphotropic high piezoelectric PMN-PT and PZN-PT compounds\cite{PhysRevB.65.064106}     & 269&  Jean-Michel Kiat & \url{10.1103/PhysRevB.65.064106}\\
4 & Continuous Variable Quantum Cryptography: Beating the 3 dB Loss Limit\cite{PhysRevLett.89.167901}   & 243  & Ch. Silberhorn&\url{10.1103/PhysRevLett.89.167901}\\
5 & Hybrid Quantum Repeater Using Bright Coherent Light\cite{PhysRevLett.96.240501}      & 209   & P. van Loock & \url{10.1103/PhysRevLett.96.240501}\\
6 & One-step deterministic polarization-entanglement purification using spatial entanglement\cite{PhysRevA.82.044305} & 204  & Yu-Bo Sheng & \url{10.1103/PhysRevA.82.044305}\\
7 & Measuring Small Longitudinal Phase Shifts: Weak Measurements or Standard Interferometry?\cite{PhysRevLett.105.010405}  & 199 & Nicolas Brunner& \url{10.1103/PhysRevLett.105.010405} \\
8 & Deterministic entanglement of superconducting qubits by parity measurement and feedback\cite{2013Deterministic} & 198 &  D. Riste& \url{10.1038/nature12513}\\
9 & Nonperturbative theory of weak pre- and post-selected measurements\cite{KOFMAN201243} & 192&Abraham G.Kofman & \url{10.1016/j.physrep.2012.07.001}\\
10 &Complex weak values in quantum measurement\cite{PhysRevA.76.044103}  & 191 & Richard Jozsa& \url{10.1103/PhysRevA.76.044103}\\

\toprule\\
\end{tabular}
\vspace*{-0.4cm}  
\end{table*}

Table. \ref{tab:analysis-of-documents} shows the top ten citation analyses of documents on weak measurement research from 2000 to 2020. "Ultrasensitive Beam Deflection Measurement via Interferometric Weak Value Amplification\cite{PhysRevLett.102.173601}" was the most popular article with the highest citations. In their work, they reported on the use of an interferometric weak value technique to amplify very small transverse deflections of an optical beam. This paper can be used as a representative work of quantum weak measurement to amplify a detector signal. In addition, the review "Colloquium: Understanding quantum weak values: Basics and applications\cite{RevModPhys.86.307}" had concluded its application to their distinct experimental techniques: weak-value amplification\cite{PhysRevLett.126.220801,PhysRevA.103.053518,PhysRevLett.125.020405}, direct quantum state and geometric determination\cite{DErrico:17,2014Direct,0Direct} and a measurable window into non-classical features of quantum mechanics\cite{Yokota_2009,Neves_2009,Lund_2010}. 

\subsection{Analysis and Co-occurrence Network of Keywords}
We analyzed a total of 634 keywords and the top 10 relevant keywords in Fig. \ref{Fig:MostrelavantWorlds}a. In our statistics, the keyword "weak measurement" had the most frequent occurrence, followed by "weak value" and "postselection". These words are corresponding to the search term On WOS in our study. "quantum optics" may suggest that quantum weak measurement is mainly achieved by photons so far\cite{CHEN2017349,He_2014}. In addition, other keywords such as "weak value amplification" "entanglement", and "quantum metrology" present the main applications of weak measurement.

Fig. \ref{Fig:WordsCloud} shows Keywords cloud on weak measurement research from 2000 to 2020. With this visual representation of the Biblioshiny Bibliometrix R-package, words/phrases with higher volume and keyword density are displayed in a larger, more prominent font. Therefore, We can get more visualized information from the Keywords cloud.

\section{Discussion}

\subsection{General Trends in Weak Measurement Research}
In this study, we combined bibliophily for Bibliometrix analyses with network visualizations to characterize the current landscape of the weak measurement research, analyzing the contributions of countries, institutions, journals, authors, high-citation documents, and keywords to this field. Since the field emerged in 2000, the annual publication output in the field has increased steadily. With the largest number of publications and citations and the top rank for co-occurrence analysis by country, the United States is currently the world leader in weak Measurement research. Even the institution University Rochester with the most publications belongs to the United States. These results suggest that the United States may have a significant impact on the direction of research in this field. China was ranked second in the total number of publications and the total citations; it was ranked second in collaboration with other countries. The top two sources ranked by citations were Physical Review A(39.7$\%$ of total citations) and Physical Review Letter(36.1$\%$ of total citations). Andrew N. Jordan(21 articles) and John C. Howell(17 articles) were the most active authors.

We concluded the general Trends in weak measurement research from 2000 to 2020 with countries, institutions, journals, and authors in Fig. \ref{Fig:institution-source-authour}. The interconnections among countries, journals, and institutions in Fig.\ref{Fig:institution-source-authour} can provide useful insights. For instance, studies on weak measurement research are mostly published in Physical Review A, the majority of which are authored by USA scholars from the University Rochester. In addition,  Fig.\ref{Fig:institution-source-authour}b shows the interactions among the most active institutions, journals, and authors. The top active authors with high-quality papers belong to the institutions of China or the USA. And Fig.\ref{Fig:institution-source-authour}c shows the interactions among the active countries, institutions, and authors. In general, the USA and China had taken a leading position with high-citations and high-quality papers in the research of weak measurement.

\subsection{Factorial Analysis on keywords}
The Biblioshiny for Bibliometrix allows using the conceptual Structure-function to perform multiple correspondence analysis (MCA) to draw a conceptual structure of the field and K-means clustering to identify clusters of documents that express common concepts\cite{ARIA2017959}. MCA is an exploratory multivariate technique for the graphical and numerical analysis of multivariate categorical data\cite{2006Multiple}. It examines the interdependence among a set of categorical variables, aiming to identify new latent variables, i.e. factors. The results are interpreted based on the relative positions of the points and their distribution along the dimensions; as words are more similar in distribution, the closer they are represented in Fig. \ref{Fig:MCA}. 

Note that new knowledge emerged from Factorial analysis of keywords. K-means clustering, also known as unsupervised classification, aims to divide data into meaningful or useful groups (or clusters). The keywords were divided into three groups by MCA. We found that the group with "interferometer", "value. amplification" and "precision" was corresponding the main application of weak measurement on weak-value amplification\cite{PhysRevLett.126.220801,PhysRevA.103.053518,PhysRevLett.125.020405}.  

\subsection{Future Outlook}
Our co-occurrence network maps, clustered by topic area or publication date, indicated the current hot topics and future directions in weak measurement research in Fig.\ref{Fig:trendtopics}. The keywords indicated that weak measurements are involved a variety of metrology(amplification, noise, value amplification), quantum communication(security, decoherence, communication) and nonclassical features of quantum mechanics(inequality, violation). The latest keywords that indicate future trends in this field are as follows.

\textbf{1}. "Security" was the most popular topic in recent years on weak measurement research. Recently, weak measurements have been proposed as a method of assessing the security in the quantum information theory\cite{Jacob2016,Unconditional_2018,Xu_2019}. By utilizing the quantum weak measurement technology, Xu et al.\cite{Xu_2019} proposed the chain inequality violation with three parties, and the analysis result demonstrates that double Chain inequality violation can be observed in the case of Alice and Bob have two different measurement bases. To improve the security of the practical quantum key distribution system, Li et al.\cite{Unconditional_2018} proposed the weak measurement model to monitor the intercept-resend eavesdropping strategy in the quantum channel, where the detector-blinding attack and the wavelength attack can be observed through the quantum bit error rate value in the weak measurement model.

\textbf{2}. "Amplification" and "noise" were the keywords of the weak-value amplification(WVA) technique to amplify a detector signal. The weak-value-amplification technique had shown great importance in the measurement of tiny physical effects\cite{PhysRevA.100.012109,Fang:21,PhysRevLett.126.220801}. In a quantum-noise limited system, the WVA technique using postselection normally does not produce more sensitive measurements because the increased weak value will lead to a small postselection probability. However, Krafczyk et al\cite{PhysRevLett.126.220801}. experimentally demonstrated recycled weak-value measurements. By using photon-counting detectors, they demonstrated a signal improvement by a factor of $4.4 \pm 0.2$ and a signal-to-noise ratio improvement of $2.1 \pm 0.06$, compared to a single-pass weak-value experiment. In addition, Huang et al\cite{PhysRevA.100.012109} proposed an approach called dual weak-value amplification (DWVA), with sensitivity several orders of magnitude higher than the standard approach without losing signal intensity. We can conclude that achieving higher sensitivity and the signal-noise rate is an earnest purpose for the weak-value-amplification technique.

\textbf{3}. "Decoherence" in terms of quantum decoherence is caused by the interaction between the quantum system. Quantum decoherence plays a pivotal role in the dynamical description of the quantum-to-classical transition and is the main impediment to the realization of devices for quantum information processing. Recently, weak measurements, which consist of several positive operators valued measures are performed before or after the evolution of the quantum system to reduce the decoherence of the system caused by the environment\cite{DATTA2017897,PRAMANIK20133209,Enhanced2019,PhysRevA.98.042322}. Datta et al\cite{DATTA2017897} demonstrated that the technique of weak measurement can be used to slow down the process of decoherence, thereby helping to preserve the quantum secret key rate when one or both systems are interacting with the environment via an amplitude damping channel. Pramanik et al\cite{PRAMANIK20133209} employed the technique of weak measurement to enable the preservation of teleportation fidelity for two-qubit noisy channels. Gupta et al\cite{PhysRevA.98.042322} shown that the diminishing effect modeled by the amplitude damping channel can be slowed down by employing the technique of weak measurements and reversals. 

\section{Conclusion}
We have drawn scientific maps of journals, countries, documents,  institutions, authors, co-occurrence network of keywords to determine trend topics in this field. With the steady growth in global trends in publications, weak measurement has become a hot topic among scientists over the past 20 years. Our results indicate that the United States is a major contributor to weak measurement research. University Rochester, University of Science and Technology of China, and Chapman University are considered excellent institutions for donating most high-citation articles. Andrew N. Jordan is a prominent researcher in this field. The Biblioshiny for Bibliometrix analysis of the publications on weak measurement is of great significance for researchers in finding the most active authors, institutions, discovering research hot-spots, and predicting the trends of weak measurement research.

\vspace{6pt} 



\authorcontributions{Jing-Hui Huang and Xue-Ying Duan collected the literature and wrote the article. Guang-Jun Wang, Fei-Fan He and Xiang-Yun Hu revised the article. Jing-Hui Huang designed the study. Xue-Ying Duan prepared figures and tables. All the authors were involved in the preparation of the manuscript. All the authors have read and approved the final manuscript.}

\funding{This research was funded by the National Key Research and Development Program of China (Grant No. 2018YFC1503705) and the Fundamental Research Funds for National Universities, China University of Geosciences (Wuhan) (Grant No. G1323519204). }

\end{paracol}



\bibliography{reference}

\end{document}